\documentclass[12pt]{article}
\usepackage[a4paper,margin=1in]{geometry}
\usepackage[T1]{fontenc}
\usepackage{setspace}
\usepackage[authoryear,round]{natbib}
\usepackage{amsmath,amssymb}
\usepackage{booktabs,tabularx}
\usepackage{placeins}
\usepackage{hyperref}
\hypersetup{colorlinks=true,linkcolor=blue,citecolor=blue,urlcolor=blue}
\title{Validity, Reliability, and Transparency in Artificial Intelligence Regulation}

\author{A.~Mukundan$^{b}$ \and Debayan~Gupta$^{a,b}$ \and Subhashis~Banerjee$^{a,b}$}
\date{$^{a}$Department of Computer Science, Ashoka University\\[4pt]
$^{b}$Centre for Digitalisation, AI, and Society, Ashoka University\\[8pt]\today}
\newenvironment{validitytable}{\begin{table}[tbp]}{\end{table}}
\begin{document}
\maketitle

\begin{abstract}
AI systems increasingly produce claims about people that shape access to
employment, healthcare, and other consequential services. Yet lawful data
processing and predictive accuracy do not establish that these claims justify
the treatment that follows. We argue that regulation must therefore examine the
legitimacy of the inferential step connecting data to claims and claims to
decisions. This requires construct, internal, and external validity: evidence
must support the meaning attributed to an output, the relationship asserted,
and its application to the intended people and setting. We show why predictive
performance alone cannot provide this warrant. Non-identifiability limits what
observations can explain, while omissions and confounding require domain-specific
judgment about what the evidence supports. Building on validity provisions in
the EU AI Act and NIST AI RMF Playbook, we develop an explicit evidentiary burden
for consequential reliance. Its normative basis follows from the dignity,
autonomy, and informational-privacy principles in \emph{Puttaswamy}
\citep{Puttaswamy2017Privacy}, which we extend from scrutiny of information
acquisition to the justification of derived claims and their use. On this
account, benefits that depend on an inference can carry weight in proportionality
assessment only to the extent that the inference is substantiated. We make this
requirement operational through a claim-specific \emph{Validity Case} linking
evidence and assumptions to permitted uses, independent review, monitoring, and
remedies. Hiring and LLM applications in healthcare and legal assistance
illustrate how the framework can guide oversight of consequential public and
private services through the relevant legal instruments.

\end{abstract}

\section{Introduction}

AI systems increasingly inform access to employment, healthcare, education,
credit, welfare, and justice. Institutions use their outputs to characterise
people and justify decisions that shape their opportunities. Lawful collection
and secure processing do not establish that these characterisations are always warranted.
\emph{Validity concerns the legitimacy of inference, not merely predictive
performance.} A correct prediction can rest on inadequate grounds for the
interpretation attached to it. Conversely, a warranted probabilistic forecast can
be followed by an unlikely outcome. Validity therefore depends on the grounds
for the inference, rather than on predictive accuracy scores. The question is what the evidence entitles an institution
to claim and how much reliance that claim can bear.

Consider a hiring system trained to predict how long employees remain in a job.
An employer may apply the resulting population regularity to a particular applicant,
interpret a low retention forecast as evidence of poor diligence, and reject the
application. These are different steps: applying evidence to an individual,
attributing a personal quality, and choosing an institutional action. Accurate
retention prediction does not establish that diligence has been measured or that
rejection is justified. Each step requires its own account of what the evidence
supports.

The epistemology of machine learning and measurement theory provide foundations
for examining these transitions
\citep{GroteGeninSullivan2024,CronbachMeehl1955,JacobsWallach2021}.
Critical studies identify the institutional and normative choices embedded in
computational inference \citep{Selbst2019,BarocasSelbst2016}. Building on related
concerns, the proposed right
to reasonable inferences connects their justification to relevance, acceptability,
and reliability \citep{WachterMittelstadt2019}. Existing governance addresses data, models, decisions, and aspects of
validity. We argue that the legitimacy of the inferential step connecting
them should become an explicit, claim-specific burden of justification.
\paragraph{Statement of Contributions.}
We connect these foundations to a constitutional argument and a regulatory
procedure that specifies what follows when evidence fails to support the claim
actually used. The contribution combines a claim-specific account of validity,
a derivation from informational privacy, and a Validity Case linking evidence
to permitted reliance and remedies. 

Epistemic risk---the risk that evidence does not adequately support a claim on
which a decision relies---must be assessed alongside the severity of its
consequences. These dimensions answer different questions: how well supported is
the inference, and what happens if an institution relies on it?\footnote{This differs from, but can contribute to, risks
to collective knowledge formation discussed by \citet{YangCasperBengio2026}.}
This analysis focuses on publicly or privately operated services that
materially affect people's rights or life opportunities.
 The same need for justification applies across
these services, although their legal obligations need not be identical.

\section{Existing Governance and the Justification of Inference}
\label{sec:governance}

Data protection concerns use as well as secrecy. The GDPR includes fairness,
accuracy, purpose limitation, data minimisation, and protections concerning
specified consequential automated decisions \citep[Articles~5 and~22]{GDPR2016}.
These requirements are relevant to inference, but an accurate input record can
still support an unwarranted interpretation. Protection depends partly on how
such duties reach the derived claim and its role in a decision.

The EU AI Act contains substantive requirements as well as risk classifications.
For high-risk systems, Articles~9--10 cover risk management, testing, data governance, measurement
assumptions, bias, and contextual data quality. Complementing these requirements,
Articles~13--15 cover information
for deployers, human oversight, accuracy, and robustness.
Article~10(2)(d) specifically addresses assumptions about what data measure and
represent. The Act also connects limits on inference to limits on use. 
Recital~44 identifies scientific limitations of AI systems that infer
emotions from biometric data. Article~5(1)(f) prohibits the use of AI
systems to infer emotions in workplaces and educational institutions,
within the Act's biometric scope, except for medical or safety reasons
\citep{EUAIAct2024}.

NIST AI RMF~1.0 likewise makes validity and reliability a foundation of
trustworthiness. Within that framework, MEASURE~2.5 addresses their demonstration
before deployment
\citep{NISTAIRMF2023}. The Playbook expressly distinguishes \emph{construct},
\emph{internal}, and \emph{external} validity, discusses proxies for attributes
such as hireability and honesty, and identifies confounding and spurious
correlations as validation concerns \citep[MEASURE~2.5]{NISTPlaybookMeasure}.
It also recommends documenting assumptions, consulting domain experts, and
responding to use outside validated conditions. These recommendations provide voluntary guidance, while the EU Act establishes
legal duties within its scope.

The Validity Case  presented in this paper organises these requirements around the particular claim used
to justify a decision. It asks an institution to state that claim, supply the
supporting evidence, and explain the limits of its application. It also specifies
what should follow when the justification fails: a narrower permitted use,
additional evidence, withdrawal of the claim, or reconsideration of affected
decisions. These steps give validity requirements a practical role within
existing assessment processes.

\section{Validity and Reliability of AI Inference}
\label{sec:validity}

Inferential validity concerns whether evidence and defensible assumptions
support a specified claim for its intended use. \emph{Construct validity}
asks whether the measurements and model outputs support the meaning
attributed to them: does the system assess the property it claims to assess?
\emph{Internal validity}, used here in a broader inferential sense than its
conventional causal usage, asks whether the evidence and study design warrant
the relationship or interpretation asserted within the setting studied,
including scrutiny of confounding, selection, leakage, and measurement
artefacts. \emph{External validity} asks whether that evidential warrant
extends to the populations, circumstances, time periods, and workflows in
which the system will be used. Predictive accuracy and reliability provide
relevant evidence, but cannot substitute for these three assessments.

\subsection{From Population Prediction to Individual Inference}
\label{sec:individual}

A hiring model may accurately predict recorded performance or retention from
educational background and employment history. An employer may nevertheless
interpret its output as evidence that an applicant lacks diligence, integrity,
resilience, or curiosity. \emph{This makes an additional epistemic leap: a regularity
among other people is being made to carry a characterisation of this person}.
The distinction matters because retention reflects working conditions and outside
opportunities, while performance ratings reflect managerial judgment. Prediction of either does not independently
establish personal character.

Population evidence can legitimately inform an individual forecast, but aggregate predictive success alone is insufficient warrant for consequential
individual judgment. The institution must
justify the relevance of the reference population, the information on which
comparison rests, and the interpretation of the result. \emph{An applicant can be well
represented on recorded features yet differ materially in unrecorded circumstances}.
 Applicability
is an evidential question that aggregate accuracy cannot answer on its own.

The ethical concern is equally substantive. When a person's opportunities depend
on the histories of others, their own conduct, circumstances, and capacity to
change can become subordinate to a statistical category. Dignity requires taking
the person seriously as an individual. Respect for autonomy also requires an
effective opportunity
to challenge the characterisation and present relevant evidence. These interests impose a burden beyond predictive performance: justify why
this evidence is an appropriate basis for this treatment
\citep{WachterMittelstadt2019}. Review must be accessible without requiring the
person to prove an unknowable ``true'' latent trait.

These requirements explain why construct validity remains relevant in predictive
applications. Even a narrow event forecast requires justified measurement and
applicability. Replacing an AI score with interviews or work samples does not
remove this burden, because those assessments also need validation. Where legitimate information cannot support
a claim, institutions must narrow it or change the decision procedure.

\subsection{Construct Validity and Measurement}
\label{sec:construct}

Construct validation asks whether observations support their intended
interpretation \citep{CronbachMeehl1955,JacobsWallach2021}. Competence at a
defined task and undifferentiated job suitability are different targets. Tests
must cover the relevant competence, distinguish it from neighbouring influences
such as familiarity, and support the comparisons made across groups or contexts.
Several indicators may agree because they share the same institutional bias or
measurement convention. Their agreement supports validity only when there is a
substantive account of why they measure the intended property. A composite index likewise requires a defensible
choice and weighting of its components. Those choices define what the index
represents, so statistical fit cannot settle the normative definition.

Examples include expenditure interpreted as health need despite unequal access
to care \citep{Obermeyer2019}, or toxicity labels affected by dialect and
annotation practice \citep{sap2019risk}. Section~\ref{sec:llmapplications}
examines how these validity requirements apply to LLM-generated summaries,
diagnoses, patient conversations, educational outcomes, and legal assistance.

Construct validity can be established when the target is clearly specified
and there is a defensible connection between that target and the observations.
Detection of referable diabetic retinopathy from fundus photographs provides
a useful example. The target is a clinically defined condition with visible
retinal manifestations, giving image-based assessment a substantive biological
basis. \citet{Gulshan2016} demonstrated high sensitivity and specificity against
an ophthalmologist-derived reference standard in retrospective evaluation.
The significance is not accuracy alone, but its combination with a clinically
grounded account of what the images and reference labels represent.

This evidence supports the specific task of detecting referable diabetic
retinopathy from fundus photographs. Whether the system can guide referral
or treatment decisions in a particular clinical setting requires further
validation. Disagreement among graders, image quality, missing clinical
context, and differences between patient populations still require assessment.
Multi-reader grading, clinical integration, and prospective evaluation can
address these concerns, but their adequacy must be demonstrated in the intended
setting. The regulatory implication is to permit reliance within the scope
supported by that evidence. The framework therefore specifies conditions for
justified AI inference: validity must be demonstrated for the claim and use
at issue, rather than inferred from predictive performance alone.

\subsubsection{Identification and the Limits of Explainability}
\label{sec:nonidentifiable}

\emph{If  incompatible explanations imply the same observations, those
observations cannot establish which explanation is true.} Let $O$ contain all
observed variables, including recorded outcomes, and $Z$ a latent attribute or
mechanism. Here $p(o)$ describes the distribution of recorded observations, while
$q(z\mid o)$ describes how a proposed model relates those observations to the
unobserved attribute. Without restrictions on that relationship, each admissible
$q$ defines a joint distribution with the same observed marginal:
\begin{equation}
p_q(o,z)=p(o)q(z\mid o),\qquad \int p_q(o,z)\,dz=p(o).
\label{eq:identification}
\end{equation}
For discrete $Z$, integration is replaced by summation. With at least two latent
states and unrestricted conditionals, infinitely many such joint distributions
are possible, potentially assigning incompatible meanings to the same
observations. Even exact knowledge of $p(o)$ cannot select among them: the models disagree
about what is hidden while agreeing about everything the observations reveal.
This is a limitation of what the evidence can establish, rather than a shortage
of training examples. The regulatory implication is that accuracy against observed outcomes cannot certify
a latent explanation. Institutions must therefore state and justify the
assumptions or additional evidence that exclude alternatives \citep{Manski2007,Pearl2009}.

Non-identifiability is target- and assumption-specific: restricted models may
identify some quantities while leaving others unresolved. If two admissible
accounts agree on all observable evidence but disagree on the explanatory target,
no procedure using that evidence alone can be guaranteed to select the true target
across both. More observations of the same kind cannot resolve this ambiguity.
An identifiable forecast can therefore coexist with a non-identifiable explanation,
even when its computation is transparent.

This is especially consequential when modelling itself is the objective.
Epidemiological case counts may reflect different combinations of actual infection
and the proportion of infections detected. Observed economic choices may likewise
reflect preferences, constraints on available options, or both. Models that fit
the recorded outcomes equally well can therefore give different accounts of the
underlying process and support different interventions.

Resolution requires justified restrictions or new evidence from measurement,
external information, or suitable study designs. A prior or regulariser may select
a model without identifying it from the observations. Domain analysis must justify
the restrictions and the link between new evidence and the target. Where the evidence supports only a range of possible values, rather than a single
identified value, the regulatory question is whether the proposed action remains
justified throughout that range. Uncertainty need not prevent every use, but it
must constrain uses whose justification changes across the remaining possibilities \citep{Manski2007}.

\subsubsection{Labels and Selective Observation}
\label{sec:labels}

Validation must explain both what labels mean and whose outcomes are observed.
For example, arrest records reflect offending, surveillance, detection, and
recording. Agreement with these records therefore does not independently
establish dangerousness. Employment
ratings reflect tasks, supervision, and evaluation practices as well as employee
performance. Moreover, performance is observed primarily for those hired.
Evaluation among past employees does not automatically support inference for
previously excluded applicants. These are distinct problems. A record may be inaccurate, it may accurately record
something other than the intended construct, or it may cover only a selected
part of the population. Improving accuracy against the available labels does
not resolve the other two problems. The institution must justify any extension beyond
the observed population and record where the necessary evidence is absent.

\subsection{Internal Validity, Confounding, and Omitted Variables}
\label{sec:internal}

While construct validity asks what an output means, internal validity asks whether
the evidence warrants the relationship or interpretation asserted. Conventionally
central to causal studies, internal validity is used here more broadly to ask whether the studied
relationships support the interpretation asserted within that setting. This
includes scrutiny of confounding, selection, leakage, and institutional artefacts.
It does not require every predictive feature to cause the outcome: a symptom can
be an effect of disease and still support diagnosis.

A confounder can induce an association without the causal relationship attributed
to it \citep{Pearl2009}. Educational pedigree may predict employee ratings because
both reflect prior advantage and access to favourable assignments, without
establishing diligence. The interpretive defect exists even before any future
performance collapse.

This interpretive limitation does not make every such association useless. An
association influenced by a common cause can still support useful forecasts
within a specified setting. The validity problem arises when an institution treats
that forecasting ability as evidence of a mechanism or stable personal quality
that the association does not establish. Causal identification is
required for causal explanations and interventions, not universally for prediction.
A further distinction concerns deployment itself. A predictor may estimate
outcomes accurately without showing that using its scores to allocate opportunities
will improve those outcomes. That second claim concerns the effect of a decision
policy, including how people and institutions respond to it.

Shortcut learning illustrates the same need for substantive evaluation.
Cross-site variation in radiograph prediction and annotation artefacts in
language tasks show that benchmark success can depend on contextual cues that
do not support the intended interpretation
\citep{zech2018variable,gururangan2018annotation,geirhos2020shortcut}.
Evaluation should test plausible shortcuts, leakage, and dependence on the setting.
Randomly dividing one dataset into training and test samples can preserve the
same artefact in both, allowing a model to perform well without demonstrating
the competence attributed to it.

\subsubsection{Omitted Variables and Domain-Dependent Judgment}
\label{sec:ovb}

In applications concerning human beings, observations capture only a small portion
of the biological, psychological, and social processes that may bear on an outcome.
\emph{The adequacy of this partial representation requires domain-specific
justification.} Determining whether opportunity, supervision, accessibility, or
task allocation has been omitted from a hiring assessment requires knowledge of
the work and the target. Optimisation over recorded variables cannot establish
that the necessary information has been included.

A standard calculation illustrates the statistical consequence of an omission.
Suppose
\[
Y=\alpha+\beta_1X_1+\beta_2X_2+\varepsilon,
\qquad E[\varepsilon\mid X_1,X_2]=0,
\]
with finite second moments and $\operatorname{Var}(X_1)>0$. The population
least-squares slope $b_1$ from regressing $Y$ on $X_1$ with an intercept satisfies
\begin{equation}
b_1-\beta_1
=
\beta_2\frac{\operatorname{Cov}(X_1,X_2)}
{\operatorname{Var}(X_1)}.
\label{eq:ovb}
\end{equation}
The formula shows how omitting $X_2$ changes the population regression
slope on $X_1$. When $X_2$ contributes to the outcome and covaries with
$X_1$, the slope in the model omitting $X_2$ incorporates part of that
omitted relationship. If both $\beta_2$ and
$\operatorname{Cov}(X_1,X_2)$ are positive, this slope is larger than
$\beta_1$, the coefficient in the fuller model. The distinction matters even when the model predicts accurately:
the fitted association does not by itself justify the interpretation
attached to its coefficient.  Regulators must require
an account of what the association supports and how plausible omissions affect
that interpretation
\citep{Pearl2009,ImbensRubin2015}.

Consider an institution that uses standardised test scores ($X_1$) to predict
university performance and interprets the prediction as an assessment of
academic potential. Test scores may provide meaningful evidence of demonstrated
knowledge and reasoning, but they also reflect access to schooling, preparation,
and educational resources. Suppose schooling quality ($X_2$) is positively
associated with test scores and makes an additional positive contribution to
university performance in the stipulated model. Omitting it then increases the
population regression coefficient on test scores relative to the fuller model.
The fitted association incorporates differences in educational opportunity
alongside the information conveyed by the test itself.

The consequential concern becomes clear for a student from an under-resourced
school. A lower score may partly reflect fewer opportunities to acquire the
knowledge and preparation being tested. If the institution interprets that
score as evidence of limited capacity to learn, it risks converting an
educational disadvantage into an adverse characterisation of the student.
Using that characterisation to deny admission can compound the original
disadvantage by withholding the opportunity through which the student's
capabilities might develop.

The construct-validity failure arises when a prediction of performance
under unequal educational conditions is treated as a measure of the
student's capacity to learn. Justifying that interpretation requires
substantive knowledge of schooling, preparation, and the opportunities
available to the student. Predictive optimisation alone cannot supply
that justification. 

Because omissions can affect measurement, interpretation, and applicability, they
can undermine all three forms of validity. Omitted-variable bias can also overlap
with confounding. Addressing these problems may require additional measurements,
suitable designs, proxies, structural
restrictions, or sensitivity analysis, each supported by justified assumptions.
Simply adding more variables is not a general solution. Conditioning on a
variable that is a common effect of two others---a collider---can itself introduce
bias into a causal analysis \citep{Pearl2009}. Additional personal data also require
an independent justification for their collection and use.
Institutions must identify plausible omissions, assess their implications, and
narrow or withdraw claims that legitimate evidence cannot sustain.

\subsection{External Validity and Deployment Distributions}
\label{sec:external}

External validity requires a connection between the setting in which a system
was evaluated and the setting in which it will be used. That connection may fail
when the distribution of inputs or their relationship to outcomes changes.
Such distribution shift can arise from demographic change, policy adaptation,
or measurement practices \citep{QuinoneroCandela2009,Ovadia2019}. Examples include
ImageNet performance degradation on newly collected images
\citep{recht2019imagenet}, temporal language change \citep{lazaridou2021mind},
and LLM sensitivity to prompts and evaluation settings \citep{liang2022holistic}.
Predictive uncertainty estimates can also fail under shift \citep{Ovadia2019}.

For open-ended LLM interaction, defining a probability measure over all possible inputs does not establish an empirically warranted deployment model. 
Conversation history alone cannot identify the process governing subsequent user inputs.  Inputs
may depend on user choices, model responses, interface design, and workflow.
Specifying a mathematical distribution does not establish that future interactions
will follow it. The empirical question is which population or interaction process
the evaluation represents, and why that process is appropriate for the intended
use.
Probabilistic performance claims hence require substantive assumptions connecting evaluation data to future interactions. 
Where those assumptions cannot be justified, benchmark averages lack a defensible interpretation as guarantees for the intended use.

\emph{Representativeness also requires domain-specific justification.}
Dataset inspection, size, or demographic balance cannot establish it alone.
\citet{zech2018variable} found variable cross-hospital performance in
pneumonia detection and evidence that models could exploit
hospital-specific information. Establishing transferability therefore
requires analysis of differences in patient populations, disease
prevalence, sampling processes, and measurement practices.

A bounded workflow may permit a defensible distributional model. Elsewhere,
scenario testing and analysis over plausible environments support narrower
claims. Regulators should require explicit boundaries covering tasks, populations,
languages, interfaces, tools, and model versions, with evidence connecting tests
to those boundaries. Monitoring must address changes in both the environment and
the decision policy, including feedback that changes who receives opportunities
and whose outcomes enter subsequent training.

\subsection{LLM Applications: From Generated Text to Consequential Use}
\label{sec:llmapplications}

LLMs learn to generate contextually plausible language, with additional
training often intended to improve accuracy and helpfulness. These
objectives do not ensure that a fluent response is factually correct
or that its conclusions follow from the available evidence. Applications
in which correctness is consequential therefore require explicit
validation of the claims generated and their suitability for the
intended use. The following examples illustrate this requirement.

\paragraph{Clinical summarisation.}
A clinical summary is often treated as a faithful representation of the
patient's record. Yet summarisation involves choices about which information
matters and how its evidential status is expressed. Converting ``possible
drug reaction'' into ``drug allergy'' turns a tentative observation into an
established condition. Omitting a pending investigation can make an unresolved
question appear settled, while presenting an old medication as current can
misrepresent the patient's treatment. These are changes in clinical meaning,
even when the summary remains fluent and largely factually accurate.

Evidence on clinical summarisation shows why overall quality and
preservation of clinically relevant information require separate
assessment. In an evaluation of 100 emergency-department encounters,
\citet{Williams2025EDSummaries} found hallucinations in 42\% of GPT-4
summaries and clinically relevant omissions in 47\%, although average
assessed potential harm was low. In a separate inpatient study,
\citet{Williams2025HospitalSummaries} found that GPT-4 Turbo summaries
were more concise and coherent than physician summaries but less
comprehensive and contained more errors. Together with the minor problems in the largely favourable
findings of \citet{VanVeen2024}, these results support validation for
the particular task and workflow, including explicit assessment of
omissions and changes in clinical meaning.

\paragraph{Automated diagnosis and treatment advice.}
Diagnostic advice requires more than matching a description to a plausible
condition. Its justification depends on whether the necessary history has
been obtained, relevant alternatives considered, and uncertainty recognised.
A recommendation that depends on medication history, when that history was
never elicited, lacks a necessary premise. An articulate explanation cannot
supply the missing information. Likewise, success on cases that provide all
relevant findings does not establish competence in deciding what to ask or
investigate in an incomplete clinical encounter.

\citet{Hager2024} found limitations in information gathering, diagnostic and
treatment decisions, and robustness in the models and cases evaluated.
These findings illustrate why assessment must examine the process through
which a recommendation is reached, as well as the final answer. Validation
should test history-taking, investigation, responses to missing or
conflicting information, and escalation to clinical care. Before autonomous
reliance is justified, evidence must also address prospective performance
in the intended workflow. Evidence for answering a clinical question does
not by itself justify responsibility for managing the encounter.

\paragraph{Patient-facing chatbots.}
When patients consult a chatbot, its responses can influence whether they
seek care, how urgently they act, and which symptoms they regard as
important. The relevant standard is therefore whether the interaction
supports appropriate decisions in the patient's circumstances.

Evidence from safety evaluations reveals failures at this transition.
In a physician-led red-teaming evaluation, \citet{Draelos2026} assessed
888 responses from four publicly available chatbots to 222 patient-posed
medical questions. Depending on the chatbot, between 5\% and 13\% of
responses were classified as unsafe, meaning that they could harm patients
or caregivers if acted upon. These findings demonstrate potentially
harmful advice in the tested systems, although the study does not estimate
its frequency in routine use.

Safety also depends on how people interact with the system.
In a preregistered study involving 1,298 participants and ten medical
scenarios, \citet{Bean2026} found that LLM-assisted participants identified
relevant conditions less accurately than controls using their usual
resources. Differences in choosing an appropriate course of action were
not statistically significant. Strong performance by the models when
tested alone did not translate into comparable performance with users.
The study identified failures involving incomplete information from users,
misinterpretation by models, and users not consistently following relevant
recommendations.

Together, these studies show why evaluating isolated answers is
insufficient to justify patient-facing deployment. A clinically appropriate
recommendation depends on obtaining relevant information, interpreting it
correctly, and communicating guidance that the patient can understand and
use. Validation must therefore examine complete interactions in the
intended population and setting, including recognition of uncertainty,
elicitation of missing information, and timely escalation to clinical
care. Evidence of medical knowledge alone cannot establish the validity
of autonomous patient guidance.

\paragraph{Education and learning.}
LLM tutors can help students produce correct answers without developing
the understanding those answers appear to demonstrate. In a field
experiment involving nearly a thousand high-school mathematics students,
\citet{Bastani2025} found that access to a general-purpose GPT-4 interface
improved performance during practice but led to worse performance than
the control group when assistance was withdrawn. A version designed
with tutoring safeguards largely mitigated this negative effect.
The validity problem is therefore the interpretation of assisted
performance as evidence of learning. Evaluation of educational
applications must distinguish successful task completion from
independent understanding, and assess retention and transfer to new
problems before improvements in assisted performance justify claims
of educational benefit.

\paragraph{Legal assistance and access to services.}
LLM-based legal assistance can influence whether a person pursues a
claim, applies for a benefit, or follows a particular procedure.
\citet{Dahl2024} documented fabricated legal information, acceptance of
false premises, and variation across courts, cases, and time in the
models tested. These findings expose problems of factual reliability.
A further validity question arises when a system applies general legal
information to an individual's circumstances. Even an accurate account
of a rule does not establish that the person is eligible for a benefit
or lacks a viable claim. That conclusion depends on the applicable
jurisdiction, current law, and relevant facts that the system may not
have elicited.

Validation must therefore assess both the accuracy of legal information
and the justification for applying it to the person seeking assistance.
Unsupported advice can discourage a valid claim or direct someone
towards an inappropriate procedure. Consequential recommendations
should identify their sources and factual assumptions, make unresolved
questions explicit, and provide access to qualified review.

Across these applications, validation must follow the output into its
intended use. Evidence that a system produces a plausible summary, diagnosis,
or explanation supports consequential reliance only when it also addresses
the meaning, evidential basis, and applicability of the claim being made.
Regulatory assessment should make those connections explicit and limit
reliance where they remain unsupported.

\subsection{What Randomised Evaluation Can Establish}
\label{sec:rct}

Randomised controlled trials can provide strong evidence that using an AI
system improves a specified outcome relative to a comparator. Such evidence
does not require a complete explanation of the model's internal workings.
However, randomisation addresses the attribution of effects within the
trial, rather than every aspect of inferential validity. It does not
establish that the measured endpoint adequately represents the claimed
benefit. For example, a trial showing that LLM-generated summaries reduce
documentation time establishes an efficiency benefit, but does not by
itself establish preservation of clinical meaning or safety in subsequent
care. Those claims require appropriate measures of consequential omissions,
distortions, and downstream outcomes. Choosing and validating these measures
requires domain knowledge that randomisation cannot supply.

The intervention being evaluated must also be distinguished from the
model considered in isolation. A trial showing improved outcomes with
AI-assisted diagnosis supports the combination of model, clinicians, and
workflow actually tested. Clinicians may obtain missing information,
correct recommendations, or prevent inappropriate action. The resulting
benefit does not independently validate the model's diagnostic explanations
or establish that autonomous use would be beneficial. This distinction is
reflected in CONSORT-AI's reporting requirements concerning system versions,
clinical integration, human--AI interaction, and performance errors
\citep{Liu2020CONSORTAI}. A demonstrated benefit remains relevant evidence,
but its interpretation must preserve the conditions under which it arose.

Randomisation also leaves important questions about interpretation and
application unresolved. \citet{DeatonCartwright2018} emphasise that it does
not guarantee balance in a realised trial, precision of estimation, or
applicability of an average effect to another population or a particular
individual. Trial findings acquire practical meaning through their
connection to other knowledge about the intervention and the conditions
under which it operates. For AI-assisted diagnosis, this includes whether
the observed benefit depends on specialist oversight, access to further
investigations, or clinicians correcting the system's errors. An average
benefit does not establish that every diagnostic claim produced during
the trial was warranted.

The transportability framework of \citet{PearlBareinboim2014} makes the
additional assumptions explicit. Transferring experimental findings
requires an account of relevant differences between the study and target
settings and of the causal relationships that remain invariant.
Depending on those assumptions, experimental and observational evidence
may identify the target effect, or may leave it unresolved. Domain
knowledge is therefore part of the justification for extending a trial
result. The regulatory implication is that a successful trial supports
a bounded claim about an intervention, outcome, and setting. A Validity
Case must explain both what the trial establishes and what additional
evidence and assumptions justify the proposed deployment.

\section{Fairness, Transparency, and Contestability}
\label{sec:transparency}

Validity and fairness are connected but distinct. A proxy can misrepresent the
target even in a balanced sample. Conversely, a genuine association can reflect structural
disadvantage that is an unacceptable basis for distributing opportunities
\citep{JacobsWallach2021,BarocasSelbst2016}. Statistical correction cannot decide
which criterion is legitimate. Nor is every individual forecast an ecological
fallacy: inference from group aggregates presents a specific problem, while
individual-level models still require justification of population evidence for
a particular person.

Formal fairness results identify further limits. Calibration within groups and
the two balance conditions studied by
\citet{KleinbergMullainathanRaghavan2017}---equal average scores across groups
within each outcome class---cannot generally hold together when base rates
differ, apart from special cases such as perfect prediction. Its epistemic
implication is that one satisfied metric cannot certify unqualified fairness.
Regulators must therefore require explicit justification of the criteria chosen
and the resulting burdens. \citet{HardtPriceSrebro2016} show how predictors can be adjusted to
satisfy a specified equal-opportunity criterion. Such adjustment
addresses a particular statistical requirement. Regulatory assessment
must also justify that criterion's relevance to the decision and
identify which fairness concerns remain unresolved.

For consequential AI, transparency must extend beyond how a model
produces an output to the evidential basis for interpreting and acting
on that output.  Access to source code,
model parameters, or feature attributions can help explain how an output
was produced, but does not establish what that output legitimately tells
us about the world. A hiring model, for example, may reveal which features
lowered an applicant's score without establishing that those features
indicate a lack of integrity or justify rejection. Evaluating an
explanation therefore requires clarity about its purpose and the evidence
appropriate to that purpose
\citep{lipton2018mythos,doshi2017towards}.

We accordingly argue that regulatory \emph{transparency must make the basis
for assessing validity available for scrutiny}. This includes the meaning
of the target, the justification for its measurement, the assumptions
connecting observations to conclusions, and the evidence supporting use
in the intended population and setting. Where competing interpretations
remain compatible with the observations, disclosure of the model's
computation cannot resolve that uncertainty
(Section~\ref{sec:construct}). Transparency must expose these limits and
enable affected people and reviewers to challenge the inference and its
role in the decision.

\section{From Informational Privacy to Inferential Validity}
\label{sec:constitutional}

The constitutional argument follows the use of information through its
transformations. \emph{Puttaswamy} recognises privacy as a fundamental right and
connects it to dignity and autonomy. We derive an evidentiary duty for consequential inference from these
constitutional principles \citep{Puttaswamy2017Privacy}.

\subsection{Protected Interests and Derived Characterisations}

Chandrachud J.'s discussion identifies information's recombinant character:
outputs become inputs to further processing, generating knowledge about people
beyond what they knowingly disclosed \citep[Chandrachud J., paras.~173--174]{Puttaswamy2017Privacy}.
If privacy protects autonomy in an information-mediated environment, scrutiny
cannot stop at the original record. An inferred attribution can shape employment,
access to services, or official treatment without a fresh disclosure of confidential
data. Protection of the underlying interest should follow the characterisation
into its consequential use.

An applicant may accept that their employment history is recorded accurately
while disputing its interpretation as evidence of poor integrity. A right to
correct the original record would not answer this objection: the disagreement
concerns what the institution claims to know from that record. Protecting the
underlying interests therefore requires scrutiny of the transformation from
record to attributed property, and of how that attribution shapes treatment.

\subsection{Justification Creates an Evidentiary Burden}

The judgment subjects State interference with privacy to legality, legitimate
aim, and proportionality. Kaul J. also expressly identifies procedural safeguards
\citep[Chandrachud J., para.~180; Kaul J., para.~71]{Puttaswamy2017Privacy}.
 We argue that these requirements provide a basis for extending scrutiny
to inferential validity.
When an institution defends a measure by invoking a
claim about a person, the asserted connection between that measure and its
purpose depends on the claim's warrant. A legitimate purpose cannot supply
missing evidence that the inference means what the institution says it means.

Applied to consequential AI, this reasoning places the evidentiary burden
on the institution invoking the inference. An employer that interprets a
model's score as evidence of an applicant's integrity must support that
interpretation. If the evidence supports only a narrower prediction, such
as retention under specified conditions, the employer must justify its
decision on that narrower basis and explain why retention is a relevant
and permissible selection criterion.

Chandrachud J.'s account of the 2012 Expert Group's proposed privacy
principles includes requirements of adequacy and relevance that reinforce
this reasoning \citep[Chandrachud J., para.~184, principle~(iv)]{Puttaswamy2017Privacy}.
Their extension requires
scrutiny of three relationships: data to processing purpose, evidence to derived
claim, and claim to decision. A datum may be relevant to recruitment yet insufficient
to establish character. Even when a forecast is supported, it may still be an
inappropriate ground for refusal. Treating these relationships separately prevents lawful acquisition
from becoming a blanket justification for subsequent inference.

This evidentiary burden also determines what meaningful contestation requires. Requiring a person to disprove
an unsupported latent attribution would transfer the institution's evidentiary
burden to someone who may lack the means to meet it. Meaningful contestation must
permit challenge to the inference's assumptions and relevance, with reasons,
authoritative review, and correction. The failure of a claim to be identifiable
is a ground for limiting reliance, not for declaring the score immune to challenge.

\subsection{Proportionality and Public and Private Responsibility}
\label{sec:proportionality}

Proportionality requires an institution to connect its claimed benefits
to evidence about the proposed use. Where a benefit depends on identifying
a personal attribute, uncertainty about that inference must enter the
assessment of whether the measure is suitable and proportionate.
A benefit demonstrated independently of the disputed attribution can
still count in its favour. For example, an intervention may improve an
observable outcome without establishing the model's interpretation of
individual character. The assessment should distinguish these grounds
and give each the weight its evidence supports.

The relevant comparison is between practicable courses of action in the
intended setting. Evaluation should examine whether the proposed service
offers benefits over existing practice, whether less intrusive
alternatives could achieve them, and how errors and burdens are
distributed. It should also account for changes induced by deployment,
including who gains access to services and whose outcomes become
available for subsequent evaluation. These considerations connect the
validity analysis to the institutional consequences of using the system.

The judgment recognises both threats to privacy from non-State actors
and the State's positive obligation to protect individuals
\citep[Chandrachud J., Section~T, conclusions~3(I) and~5]{Puttaswamy2017Privacy}.
We draw on these principles to propose scrutiny of consequential
inference across publicly and privately operated services. The legal
route will depend on the institution and activity. Public-law review,
sectoral regulation, data-protection duties, and procurement conditions
can each provide mechanisms for requiring evidence, enabling challenge,
and securing remedies. The framework proposed here supplies a common
basis for that scrutiny, while leaving the applicable legal duties to
be specified for each setting.

\section{A Validity-Grounded Regulatory Framework}
\label{sec:framework}

Three duties organise the Validity Case: substantiate the claim, limit reliance
to its warrant, and provide effective challenge and correction. Assessment
integrates epistemic uncertainty and consequence severity with existing oversight.
It must also preserve independent legal constraints, which strong validation
cannot waive.

\subsection{Risk and the Validity Case}

Epistemic assessment begins by asking whether the target is clearly defined,
whether the measurements support it, and which assumptions exclude alternative
interpretations. It then examines omitted information, applicability to the
intended people and setting, and feedback caused by use of the system. Assessors
must also consider whether errors can be detected and their consequences reversed.
A domain label such as healthcare or employment helps locate these questions,
but cannot answer them for a particular system. Strong evidence on one dimension
should not conceal a serious unresolved weakness on another through averaging
into a single risk score.

A Validity Case is a reasoned account of the evidence for each consequential
claim, including claims conveyed by interfaces and decision practices rather
than documentation alone. Table~\ref{tab:case} translates the preceding analysis into questions the
institution must answer, with concrete evidence for each answer.
Responsibility follows the institution's role. Providers should supply evidence
of capabilities and known limitations, while deployers must justify local
applicability and the actual decision policy. A designated assessor should examine whether the evidence actually supports the
claimed use. Completing a documentation checklist cannot establish that connection.
For example, a provider's evidence of general capability does not answer whether
a deployer's local workflow preserves the conditions under which it was tested.
Independent assessment is warranted where consequences are severe, assumptions
are contested, or institutional incentives threaten credible review.

\begin{validitytable}
\centering
\small
\begin{tabularx}{\textwidth}{@{}>{\raggedright\arraybackslash}p{0.28\textwidth}>{\raggedright\arraybackslash}X@{}}
\toprule
\textbf{Question and basis} & \textbf{What the institution must explain and demonstrate}\\
\addlinespace
\midrule
What claim will support the decision?\newline
Sections~\ref{sec:individual}--\ref{sec:construct}
& Define the target inference, intended population, setting, time horizon,
and decision role. Establish construct validity by demonstrating that
the observations support the meaning attributed to the output.
State the assumptions and evidence needed to identify the target,
and delimit what the available data can justify.\\
\addlinespace
Do the observations support that meaning?\newline
Sections~\ref{sec:construct} (labels and measurement) and~\ref{sec:llmapplications}
& Explain how inputs and labels were produced and whose outcomes are missing.
Show why they measure the stated target: employee ratings may reflect managerial
practice, while a clinical summary must preserve source meaning, timing, and uncertainty.\\
\addlinespace
What could undermine the inference?\newline
Sections~\ref{sec:construct} (identification) and~\ref{sec:internal} (confounding and omissions)
& State the assumptions that exclude competing explanations. Examine plausible
omissions, confounding, and sensitivity to alternatives. Report unresolved
ambiguity, including when a diagnostic recommendation depends on history never obtained.\\
\addlinespace
Why does the evidence apply here?\newline
Sections~\ref{sec:individual}, \ref{sec:external}--\ref{sec:llmapplications}
& Justify the comparison population and intended workflow. Evaluate relevant
groups and settings, including actual users of a patient chatbot. Specify how
supplementary information can expose an inapplicable forecast or trigger further review.\\
\addlinespace
What reliance is justified?\newline
Section~\ref{sec:constitutional}, especially~\ref{sec:proportionality}
& Connect supported claims to the proposed action. Compare policy benefits,
harms, and feasible alternatives. State use limits: evidence for drafting a
clinician-reviewed message does not establish safe autonomous advice.\\
\addlinespace
Who must answer a challenge?\newline
Sections~\ref{sec:transparency}, \ref{sec:constitutional}
& Name the responsible provider, deployer, and assessor. Give affected people
intelligible reasons and access to review. The reviewer must be able to examine
the evidence, disregard an unsupported output, and correct the resulting decision.\\
\addlinespace
When must reliance change?\newline
Sections~\ref{sec:external}--\ref{sec:llmapplications} and operational duties in
Section~\ref{sec:determinations}
& Specify monitored failures, complaints, and changes that require revalidation,
such as a new model, patient population, or workflow. Name who can suspend use
and how affected records and decisions will be reconsidered.\\
\addlinespace
\bottomrule
\end{tabularx}
\caption{A Validity Case: questions and evidence derived from the preceding analysis.}
\label{tab:case}
\end{validitytable}

Evidence standards should be specified before testing where feasible, using domain
expertise and affected communities' knowledge of consequences. No universal accuracy
threshold establishes validity. The assessor must justify the required evidence
against the claim, severity and distribution of error, reversibility, and feasible
alternatives. A uniquely identified explanation is not always necessary for justified action.
If the evidence leaves several credible possibilities, the assessor should ask
whether the same permissible action remains justified under each. Where those
possibilities instead change the basis of an adverse decision, selecting the
most convenient interpretation cannot resolve the uncertainty.

\subsection{Determinations, Monitoring, and Remedies}
\label{sec:determinations}

Assessment proceeds from claim specification to evidentiary adequacy, comparison
of supported benefits and harms, and a reasoned determination of permitted reliance.
The process can be iterative. The seriousness of the consequences helps determine
how strong the evidence must be, while the evidence limits which benefits can
legitimately be claimed. This permits revision of a proposal without treating
validation as a one-time certificate for every subsequent use. Four findings require distinct
responses:
\begin{enumerate}
\item \textbf{The claim lacks a defensible meaning or operationalisation.}
Require reformulation or withdrawal before consequential reliance.
\item \textbf{The target is meaningful but evidence is insufficient.}
Obtain appropriate evidence, narrow the use, or withhold reliance. Uncertainty
must not be treated as proof of an adverse personal attribute.
\item \textbf{Evidence supports a bounded use.}
Permit only that use, subject to the remaining legal and proportionality
requirements, enforceable restrictions, and monitoring.
\item \textbf{Evidence contradicts the interpretation or a material assumption fails.}
Suspend affected reliance, investigate scope, and reconsider past decisions and
remedies appropriate to the defect and its consequences.
\end{enumerate}

To make these determinations credible, review intensity should rise with
evidentiary difficulty and consequence severity.
Controlled evaluation can be authorised separately, with safeguards, without
licensing disputed adverse decisions. Advisory scores require scrutiny when they
materially shape treatment.

Table~\ref{tab:case} specifies revalidation triggers and authority to suspend use.
Monitoring should examine who is selected or excluded, when staff override the
system, what happens during review, and whether failures are concentrated in
particular groups. Use restrictions must be reflected in interfaces and staff
procedures: a limitation recorded only in technical documentation may have little
effect on how an output actually influences decisions.
Systemic defects require reconsideration of similarly affected cases. Remedies may
include fresh assessment, correction of derived records, and notification of
downstream recipients, subject to the applicable legal scheme.

\subsection{Worked Application: Hiring and Individual Reliance}
\label{sec:hiring}

Suppose an employer proposes automatic rejection using a model predicting
one-year retention. Its validation shows good historical performance among
employees hired under the previous recruitment process. Managers describe a low
score as evidence of poor diligence. This hypothetical proposal contains three
claims: a retention forecast, an attribution of character, and a claim that
score-based exclusion improves recruitment.

The measurement account must define retention and distinguish voluntary departure,
dismissal, and incomplete follow-up. It must identify how past selection and
working conditions shaped the observed outcomes. Validation among hired workers
does not establish performance for all applicants. Even if applicability is
established, good retention prediction does not establish diligence. Nor does
improved prediction establish that the proposed exclusion policy is beneficial. These are separate evidentiary gaps,
not three descriptions of the same accuracy problem.

The assessor should require withdrawal of the unsupported character attribution,
including its use in reason codes and managerial guidance. Automatic exclusion
would remain unsupported until evidence addresses applicability to the proposed
applicant population and the decision policy. Any narrower use must be evaluated
on its own terms. Appropriate further evidence could include a governed prospective
evaluation, investigation of selection and workplace effects, and comparison with
a validated task-based assessment. Such evaluation must not assume that historical
hiring decisions or human ratings provide an unbiased standard.

An applicant may supply relevant evidence about a career interruption, changed
circumstances, or demonstrated task performance. This does not automatically
establish a replacement probability. Its relevance lies in exposing possible
inadequacies in the original comparison that require reconsideration. Review must be accessible, and the
institution must explain whether the evidence changes the claim's applicability.
A process available only to unusually well-resourced applicants would create
another barrier to fair treatment.

Any approval should specify the job family, population, outcome, permitted decision
role, and review procedure. Monitoring must follow both the predictor and changes
to recruitment or working conditions. Unsupported characterisation is withdrawn,
unsupported exclusion is withheld, and permitted reliance remains correctable.

\section{Implementation and Limits}

The Validity Case is a proposed regulatory instrument whose effectiveness requires
empirical evaluation. It can be incorporated into conformity assessment, sectoral
supervision, procurement, or institutional approval, with duties and enforcement
powers defined in the relevant instrument. Shared provider evidence can reduce
duplication, but cannot substitute for local justification. Reviewers need domain
competence, access to evidence, independence, and resources to examine the
justification offered. A completed template alone provides no assurance.

Evidence standards involve normative judgments, and individual circumstances may
remain inaccessible. These limits should constrain reliance without demanding
unlimited personal data. Alternative human procedures require equivalent scrutiny.
The framework should itself be evaluated. Relevant questions include whether
assessors reach consistent and reasoned determinations, whether affected people
can obtain effective remedies, and whether decisions improve compared with
existing validation and oversight. Producing more documentation is not evidence
that any of these outcomes has been achieved.

\section{Conclusion}

Consequential AI requires justification of the transition from data to claim and
from claim to treatment. Population accuracy cannot alone establish individual
character, identify an underlying mechanism, or justify a decision. Construct,
internal, and external validity require domain-specific evidence and assumptions
that statistical optimisation cannot supply by itself.

Existing EU and NIST provisions establish substantial foundations for this
inquiry. The extension developed from \emph{Puttaswamy} gives it a constitutional
rationale: protection of dignity and autonomy should reach the evidentiary basis
of derived characterisations and the consequences of relying on them. Validity is
therefore a precondition for crediting inferentially dependent benefits in
proportionality assessment. The Validity Case makes that burden operational through
explicit claims, evidence, bounded reliance, authoritative review, and remedies.
\emph{An institution must justify what it claims to know about a person before
that claim can justify how the person is treated.}

\FloatBarrier
\bibliographystyle{abbrvnat}
\bibliography{refs-clean,AI-validity-new-references}

@techreport{NISTAIRMF2023,
  author = {{National Institute of Standards and Technology}},
  title = {Artificial Intelligence Risk Management Framework ({AI RMF} 1.0)},
  institution = {National Institute of Standards and Technology},
  number = {NIST AI 100-1},
  year = {2023},
  doi = {10.6028/NIST.AI.100-1},
  url = {https://nvlpubs.nist.gov/nistpubs/ai/NIST.AI.100-1.pdf}
}

@misc{NISTPlaybookMeasure,
  author = {{National Institute of Standards and Technology}},
  title = {{AI RMF Playbook}: Measure},
  howpublished = {AI Resource Center},
  year = {n.d.},
  note = {Living guidance, MEASURE 2.5; accessed 11 September 2026},
  url = {https://airc.nist.gov/airmf-resources/playbook/measure/}
}

@misc{GDPR2016,
  author = {{European Parliament and Council of the European Union}},
  title = {Regulation ({EU}) 2016/679 ({General Data Protection Regulation})},
  year = {2016},
  howpublished = {Official Journal of the European Union, L 119, 4 May 2016},
  url = {https://eur-lex.europa.eu/eli/reg/2016/679/oj}
}

@article{VanVeen2024,
  author = {Van Veen, Dave and Van Uden, Cara and Blankemeier, Louis and others},
  title = {Adapted Large Language Models Can Outperform Medical Experts in Clinical Text Summarization},
  journal = {Nature Medicine},
  volume = {30},
  pages = {1134--1142},
  year = {2024},
  doi = {10.1038/s41591-024-02855-5},
  url = {https://www.nature.com/articles/s41591-024-02855-5}
}

@article{Hager2024,
  author = {Hager, Paul and Jungmann, Friederike and Holland, Robbie and Bhagat, Kunal and Hubrecht, Inga and Knauer, Manuel and Vielhauer, Jakob and Makowski, Marcus and Braren, Rickmer and Kaissis, Georgios and Rueckert, Daniel},
  title = {Evaluation and Mitigation of the Limitations of Large Language Models in Clinical Decision-Making},
  journal = {Nature Medicine},
  volume = {30},
  number = {9},
  pages = {2613--2622},
  year = {2024},
  doi = {10.1038/s41591-024-03097-1},
  url = {https://www.nature.com/articles/s41591-024-03097-1}
}

@article{Bean2026,
  author = {Bean, Andrew M. and Payne, Rebecca Elizabeth and Parsons, Guy and others},
  title = {Reliability of {LLMs} as Medical Assistants for the General Public: A Randomized Preregistered Study},
  journal = {Nature Medicine},
  volume = {32},
  number = {2},
  pages = {609--615},
  year = {2026},
  doi = {10.1038/s41591-025-04074-y},
  url = {https://www.nature.com/articles/s41591-025-04074-y}
}

@article{Dahl2024,
  author = {Dahl, Matthew and Magesh, Varun and Suzgun, Mirac and Ho, Daniel E.},
  title = {Large Legal Fictions: Profiling Legal Hallucinations in Large Language Models},
  journal = {Journal of Legal Analysis},
  volume = {16},
  number = {1},
  pages = {64--93},
  year = {2024},
  doi = {10.1093/jla/laae003},
  url = {https://doi.org/10.1093/jla/laae003}
}

@article{Liu2020CONSORTAI,
  author  = {Liu, Xiaoxuan and Cruz Rivera, Samantha
             and Moher, David and Calvert, Melanie J.
             and Denniston, Alastair K.
             and {SPIRIT-AI and CONSORT-AI Working Group}},
  title   = {Reporting Guidelines for Clinical Trial Reports
             for Interventions Involving Artificial Intelligence:
             The {CONSORT-AI} Extension},
  journal = {BMJ},
  year    = {2020},
  volume  = {370},
  pages   = {m3164},
  doi     = {10.1136/bmj.m3164}
}

@article{DeatonCartwright2018,
  author  = {Deaton, Angus and Cartwright, Nancy},
  title   = {Understanding and Misunderstanding Randomized
             Controlled Trials},
  journal = {Social Science \& Medicine},
  year    = {2018},
  volume  = {210},
  pages   = {2--21},
  doi     = {10.1016/j.socscimed.2017.12.005}
}

@article{PearlBareinboim2014,
  author  = {Pearl, Judea and Bareinboim, Elias},
  title   = {External Validity: From Do-Calculus to
             Transportability Across Populations},
  journal = {Statistical Science},
  year    = {2014},
  volume  = {29},
  number  = {4},
  pages   = {579--595},
  doi     = {10.1214/14-STS486}
}

@article{Williams2025EDSummaries,
  author  = {Williams, Christopher Y. K. and Bains, Jaskaran
             and Tang, Tianyu and Patel, Kishan
             and Lucas, Alexa N. and Chen, Fiona
             and Miao, Brenda Y. and Butte, Atul J.
             and Kornblith, Aaron E.},
  title   = {Evaluating Large Language Models for Drafting
             Emergency Department Encounter Summaries},
  journal = {PLOS Digital Health},
  year    = {2025},
  volume  = {4},
  number  = {6},
  pages   = {e0000899},
  doi     = {10.1371/journal.pdig.0000899}
}

@article{Williams2025HospitalSummaries,
  author  = {Williams, Christopher Y. K.
             and Subramanian, Charumathi Raghu
             and Ali, Syed Salman and Apolinario, Michael
             and Askin, Elisabeth and Barish, Peter
             and Cheng, Monica and Deardorff, W. James
             and Donthi, Nisha and Ganeshan, Smitha
             and Huang, Owen and Kantor, Molly A.
             and Lai, Andrew R. and Manchanda, Ashley
             and Moore, Kendra A. and Muniyappa, Anoop N.
             and Nair, Geethu and Patel, Prashant P.
             and Santhosh, Lekshmi and Schneider, Susan
             and Torres, Shawn and Yukawa, Michi
             and Hubbard, Colin C. and Rosner, Benjamin I.},
  title   = {Physician- and Large Language Model-Generated
             Hospital Discharge Summaries},
  journal = {JAMA Internal Medicine},
  year    = {2025},
  volume  = {185},
  number  = {7},
  pages   = {818--825},
  doi     = {10.1001/jamainternmed.2025.0821}
}

@article{Bastani2025,
  author  = {Bastani, Hamsa and Bastani, Osbert and Sungu, Alp
             and Ge, Haosen and Kabakc{\i}, {\"O}zge
             and Mariman, Rei},
  title   = {Generative {AI} without guardrails can harm learning:
             Evidence from high school mathematics},
  journal = {Proceedings of the National Academy of Sciences},
  year    = {2025},
  volume  = {122},
  number  = {26},
  pages   = {e2422633122},
  doi     = {10.1073/pnas.2422633122}
}

@article{Gulshan2016,
  title     = {Development and Validation of a Deep Learning Algorithm for Detection of Diabetic Retinopathy in Retinal Fundus Photographs},
  author    = {Gulshan, Varun and Peng, Lily and Coram, Marc and Stumpe, Martin C. and Wu, Derek and Narayanaswamy, Arunachalam and Venugopalan, Subhashini and Widner, Kasumi and Madams, Tom and Cuadros, Jorge and others},
  journal   = {{JAMA}},
  volume    = {316},
  number    = {22},
  pages     = {2402--2410},
  year      = {2016},
  doi       = {10.1001/jama.2016.17216}
}

@article{GroteGeninSullivan2024,
  author    = {Grote, Thomas and Genin, Konstantin and Sullivan, Emily},
  title     = {Reliability in Machine Learning},
  journal   = {Philosophy Compass},
  year      = {2024},
  volume    = {19},
  number    = {5},
  pages     = {e12974},
  doi       = {10.1111/phc3.12974},
  url       = {https://doi.org/10.1111/phc3.12974}
}

@book{QuinoneroCandela2009,
  editor    = {Qui{\~n}onero-Candela, Joaquin and Sugiyama, Masashi and Schwaighofer, Anton and Lawrence, Neil D.},
  title     = {Dataset Shift in Machine Learning},
  publisher = {MIT Press},
  year      = {2009},
  url       = {https://dl.acm.org/doi/10.5555/1462129},
  note      = {ACM record for the MIT Press volume}
}

@inproceedings{Ovadia2019,
  author    = {Ovadia, Yaniv and Fertig, Emily and Ren, Jie and Nado, Zachary and Sculley, D. and Nowozin, Sebastian and Dillon, Joshua V. and Lakshminarayanan, Balaji and Snoek, Jasper},
  title     = {Can You Trust Your Model's Uncertainty? Evaluating Predictive Uncertainty Under Dataset Shift},
  booktitle = {Advances in Neural Information Processing Systems},
  volume    = {32},
  year      = {2019},
  url       = {https://arxiv.org/abs/1906.02530}
}

@article{CronbachMeehl1955,
  author    = {Cronbach, Lee J. and Meehl, Paul E.},
  title     = {Construct Validity in Psychological Tests},
  journal   = {Psychological Bulletin},
  year      = {1955},
  volume    = {52},
  number    = {4},
  pages     = {281--302},
  doi       = {10.1037/h0040957},
  url       = {https://doi.org/10.1037/h0040957}
}

@book{Pearl2009,
  author = {Pearl, Judea},
  title = {Causality: Models, Reasoning, and Inference},
  edition = {2},
  publisher = {Cambridge University Press},
  year = {2009},
  url = {https://bayes.cs.ucla.edu/BOOK-09/causality2-excerpts.htm}
}

@article{BarocasSelbst2016,
  author    = {Barocas, Solon and Selbst, Andrew D.},
  title     = {Big Data's Disparate Impact},
  journal   = {California Law Review},
  year      = {2016},
  volume    = {104},
  number    = {3},
  pages     = {671--732},
  url       = {https://www.cs.yale.edu/homes/jf/BarocasSelbst.pdf}
}

@inproceedings{HardtPriceSrebro2016,
  author    = {Hardt, Moritz and Price, Eric and Srebro, Nati},
  title     = {Equality of Opportunity in Supervised Learning},
  booktitle = {Advances in Neural Information Processing Systems},
  volume    = {29},
  year      = {2016},
  url       = {https://arxiv.org/abs/1610.02413}
}

@inproceedings{KleinbergMullainathanRaghavan2017,
  author    = {Kleinberg, Jon and Mullainathan, Sendhil and Raghavan, Manish},
  title     = {Inherent Trade-Offs in the Fair Determination of Risk Scores},
  booktitle = {Proceedings of the 8th Innovations in Theoretical Computer Science Conference (ITCS 2017)},
  year      = {2017},
  url       = {https://arxiv.org/abs/1609.05807}
}

@inproceedings{Selbst2019,
  author    = {Selbst, Andrew D. and Boyd, Danah and Friedler, Sorelle A. and Venkatasubramanian, Suresh and Vertesi, Janet},
  title     = {Fairness and Abstraction in Sociotechnical Systems},
  booktitle = {Proceedings of the Conference on Fairness, Accountability, and Transparency (FAT* '19)},
  year      = {2019},
  pages     = {59--68},
  doi       = {10.1145/3287560.3287598},
  url       = {https://doi.org/10.1145/3287560.3287598}
}

@article{zech2018variable,
  title = {Variable Generalization Performance of a Deep Learning Model to Detect Pneumonia in Chest Radiographs: A Cross-Sectional Study},
  author = {Zech, John R. and Badgeley, Marcus A. and Liu, Manway and Costa, Anthony B. and Titano, Joseph J. and Oermann, Eric Karl},
  journal = {{PLOS} Medicine},
  volume = {15},
  number = {11},
  pages = {e1002683},
  year = {2018},
  doi = {10.1371/journal.pmed.1002683},
  url = {https://journals.plos.org/plosmedicine/article?id=10.1371/journal.pmed.1002683}
}

@inproceedings{gururangan2018annotation,
  title     = {Annotation Artifacts in Natural Language Inference Data},
  author    = {Gururangan, Suchin and Swayamdipta, Swabha and Levy, Omer and Schwartz, Roy and Bowman, Samuel and Smith, Noah A.},
  booktitle = {Proceedings of the 2018 Conference of the North American Chapter of the Association for Computational Linguistics: Human Language Technologies (NAACL-HLT)},
  pages     = {107--112},
  year      = {2018}
}

@article{geirhos2020shortcut,
  title     = {Shortcut Learning in Deep Neural Networks},
  author    = {Geirhos, Robert and Jacobsen, J{\"o}rn-Henrik and Michaelis, Claudio and Zemel, Richard and Brendel, Wieland and Bethge, Matthias and Wichmann, Felix A.},
  journal   = {Nature Machine Intelligence},
  volume    = {2},
  number    = {11},
  pages     = {665--673},
  year      = {2020}
}

@inproceedings{recht2019imagenet,
  title     = {Do {ImageNet} Classifiers Generalize to {ImageNet}?},
  author    = {Recht, Benjamin and Roelofs, Rebecca and Schmidt, Ludwig and Shankar, Vaishaal},
  booktitle = {Proceedings of the 36th International Conference on Machine Learning (ICML)},
  pages     = {5389--5400},
  year      = {2019}
}

@inproceedings{lazaridou2021mind,
  title = {Mind the Gap: Assessing Temporal Generalization in Neural Language Models},
  author = {Lazaridou, Angeliki and Kuncoro, Adhiguna and Gribovskaya, Elena and Agrawal, Devang and Liska, Adam and Terzi, Tayfun and Gimenez, Mai and de Masson d'Autume, Cyprien and Kocisky, Tomas and Ruder, Sebastian and Yogatama, Dani and Cao, Kris and Young, Susannah and Blunsom, Phil},
  booktitle = {Advances in Neural Information Processing Systems},
  volume = {34},
  pages = {29348--29363},
  year = {2021},
  url = {https://arxiv.org/abs/2102.01951}
}

@article{liang2022holistic,
  title = {Holistic Evaluation of Language Models},
  author = {Liang, Percy and Bommasani, Rishi and Lee, Tony and Tsipras, Dimitris and Soylu, Dilara and Yasunaga, Michihiro and others},
  journal = {arXiv preprint arXiv:2211.09110},
  year = {2022},
  url = {https://arxiv.org/abs/2211.09110}
}

@article{lipton2018mythos,
  title     = {The Mythos of Model Interpretability},
  author    = {Lipton, Zachary C.},
  journal   = {Queue},
  volume    = {16},
  number    = {3},
  pages     = {31--61},
  year      = {2018}
}

@article{doshi2017towards,
  title     = {Towards a Rigorous Science of Interpretable Machine Learning},
  author    = {Doshi-Velez, Finale and Kim, Been},
  journal   = {arXiv preprint arXiv:1702.08608},
  year      = {2017}
}

@misc{Puttaswamy2017Privacy,
  author       = {{Supreme Court of India}},
  title        = {Justice {K.S. Puttaswamy} (Retd.) and Anr. v. Union of India and Ors., Writ Petition (Civil) No. 494 of 2012},
  howpublished = {Supreme Court of India},
  year         = {2017},
  note         = {(2017) 10 SCC 1. Nine-judge constitutional bench unanimously recognizing the right to privacy as a fundamental right under the Constitution of India}
}

@book{Manski2007,
  author    = {Manski, Charles F.},
  title     = {Identification for Prediction and Decision},
  publisher = {Harvard University Press},
  year      = {2007}
}

@article{Obermeyer2019,
  author    = {Obermeyer, Ziad and Powers, Brian and Vogeli, Christine and Mullainathan, Sendhil},
  title     = {Dissecting Racial Bias in an Algorithm Used to Manage the Health of Populations},
  journal   = {Science},
  year      = {2019},
  volume    = {366},
  number    = {6464},
  pages     = {447--453},
  doi       = {10.1126/science.aax2342}
}

@book{ImbensRubin2015,
  author    = {Imbens, Guido W. and Rubin, Donald B.},
  title     = {Causal Inference for Statistics, Social, and Biomedical Sciences: An Introduction},
  publisher = {Cambridge University Press},
  address   = {New York},
  year      = {2015},
  isbn      = {978-0-521-88588-1},
  doi       = {10.1017/CBO9781139025751},
  url       = {https://doi.org/10.1017/CBO9781139025751}
}

@inproceedings{sap2019risk,
  author    = {Sap, Maarten and Card, Dallas and Gabriel, Saadia and Choi, Yejin and Smith, Noah A.},
  title     = {The Risk of Racial Bias in Hate Speech Detection},
  booktitle = {Proceedings of the 57th Annual Meeting of the Association for Computational Linguistics},
  pages     = {1668--1678},
  year      = {2019},
  address   = {Florence, Italy},
  publisher = {Association for Computational Linguistics},
  doi       = {10.18653/v1/P19-1163}
}

@inproceedings{JacobsWallach2021,
  author    = {Jacobs, Abigail Z. and Wallach, Hanna},
  title     = {Measurement and Fairness},
  booktitle = {Proceedings of the 2021 {ACM} Conference on Fairness, Accountability, and Transparency},
  series    = {FAccT '21},
  pages     = {375--385},
  year      = {2021},
  address   = {New York, NY, USA},
  publisher = {Association for Computing Machinery},
  doi       = {10.1145/3442188.3445901},
  url       = {https://doi.org/10.1145/3442188.3445901}
}

@misc{EUAIAct2024,
  author       = {{European Parliament and Council of the European Union}},
  title        = {Regulation ({EU}) 2024/1689 of the {European Parliament} and of the {Council} of 13 {June} 2024 Laying Down Harmonised Rules on Artificial Intelligence ({Artificial Intelligence Act}) and Amending Certain {Union} Legislative Acts},
  howpublished = {Official Journal of the European Union},
  year         = {2024},
  month        = jul,
  note         = {OJ L, 2024/1689, 12.7.2024. \url{https://eur-lex.europa.eu/legal-content/EN/TXT/?uri=OJ:L_202401689}}
}

@unpublished{YangCasperBengio2026,
  author    = {Yang, Mick and Casper, Stephen and Stray, Jonathan and Li, Jasmine and Jones, Cameron and Gausen, Anna and Jaques, Natasha and Christian, Brian and Gyevn{\'a}r, B{\'a}lint and Kirk, Hannah and He, Zhonghao and Zhao, Dan and Looi, Siao Si and Levy, Joshua and Hackenburg, Kobi and Seger, Elizabeth and Kowal, Matt and Malonza, Michelle and Hewitt, Luke and Lin, Hause and Sap, Maarten and Hadfield-Menell, Dylan and Costello, Thomas and Rabbany, Reihaneh and Godbout, Jean-Fran{\c{c}}ois and Rand, David and Kasirzadeh, Atoosa and Pennycook, Gordon and Bengio, Yoshua and Pelrine, Kellin},
  title     = {{AI} Epistemic Risks: Emerging Mechanisms \& Evidence},
  note      = {SSRN preprint 6873005},
  year      = {2026},
  month     = jun,
  doi       = {10.2139/ssrn.6873005},
  url       = {https://ssrn.com/abstract=6873005}
}

@article{WachterMittelstadt2019,
  author    = {Wachter, Sandra and Mittelstadt, Brent},
  title     = {A Right to Reasonable Inferences: Re-Thinking Data Protection Law in the Age of Big Data and {AI}},
  journal   = {Columbia Business Law Review},
  volume    = {2019},
  number    = {2},
  pages     = {494--621},
  year      = {2019},
  url       = {https://ssrn.com/abstract=3248829}
}

@article{Draelos2026,
  author  = {Draelos, Rachel L. and Afreen, Samina and
             Blasko, Barbara and others},
  title   = {Large language models provide unsafe answers
             to patient-posed medical questions},
  journal = {npj Digital Medicine},
  volume  = {9},
  pages   = {241},
  year    = {2026},
  doi     = {10.1038/s41746-026-02428-5}
}
\end{document}